\documentstyle[11pt,newpasp,twoside,epsf,graphicx]{article}

\makeindex 
\index{a:Amram, P.}
\index{a:Bournaud, F.}
\index{a:Duc, P.--A.}

\index{k:Interactions!numerical simulations}
\index{k:Interactions!collisions}
\index{k:Intergalactic medium!intergalactic HII regions}
\index{k:Tidal Dwarf Galaxies!kinematics}
\index{k:Tidal Dwarf Galaxies!simulations}
\index{k:Tidal Dwarf Galaxies!dark matter}

\index{o:IC 1182}
\index{o:Arp 105 (NGC 3561)}
\index{o:Arp 242 (NGC 4676)}
\index{o:NGC 5291}

\def\edcomment#1{\iffalse\marginpar{\raggedright\sl#1\/}\else\relax\fi}
\reversemarginpar

\begin{document}
\title{$H\alpha$ Kinematics of Tidal Tails in Interacting Systems:  Projection
 Effects and  Dark Matter in TDGs}
\author{Philippe Amram$^1$, Fr\'{e}deric Bournaud$^{2,3}$,  Pierre--Alain Duc$^2$} 
\affil{$^1$ Observatoire de Marseille, France\\
$^2$ CEA, Saclay, France\\
$^3$ Observatoire de Paris, France}

\begin{abstract}
\noindent 
Several interacting systems exhibit at the tip of their long tidal tails
 massive condensations of atomic hydrogen, which may be the progenitors of
Tidal Dwarf Galaxies. Because, quite often, these tails
are observed edge-on, projection effects have been claimed to
account for the large HI column densities measured there. Here we show that 
determining the velocity field all along the tidal features, one may disentangle 
projection effects along the line of view from real bound structures. Due
to its large field of view, high spectral and 2D spatial resolutions,
Fabry-Perot observations of the ionized gas are well adapted to 
detect a kinematical signature of either streaming motions along a 
bent tidal tail or of infalling/rotating material associated with a forming TDG.
Spectroscopic observations also allow to measure the
dynamical masses of the TDGs that are already relaxed and
check their dark matter content. 
\end{abstract}

\section{Introduction: the kinematics of tidal tails}
The most impressive and surely most studied interacting systems,
such as the Antennae galaxies, exhibit long optical tidal tails that may extend 
up to 100 kpc. 
HI observations of such colliding galaxies have shown that the stellar tails have
a gaseous counterpart that is usually even more prominent and contain
a large fraction of the total atomic gas present in the system.
In a few interacting galaxies, the HI tidal tails exhibit some gas concentrations
that have apparent masses of up to few $10^{9}$ solar masses. Those
are the progenitors of the 
Tidal dwarf galaxies (TDGs), at least the more massive ones.
They are typically found at the tip of the optical
tidal tails at distances between 30 and 100 kpc from the merging
disks. They might be as massive as the Magellanic Clouds, and on top
of being rich in atomic hydrogen gas (Duc et al. 2000), they contain high
quantities of molecular gas (Braine et al. 2001, and his contribution in this volume) and form stars 
with a rate as high as in blue compact dwarf galaxies (Duc \& Mirabel 1998). 
The amount of stellar and gaseous material in several TDGs suggests that they are 
gravitationally bound, although we do not have yet conclusive observational evidence
 of such a self-gravitating TDG, kinematically independent from
its host tidal tail. Moreover if dark matter is made of collisionless material distributed in a
large halo, it should not be ejected with tidal stellar and
gaseous debris pulled out from colliding disks (Barnes \&
Hernquist, 1992). In that case, TDGs should not
contain a significant amount of DM whereas ordinary dwarf galaxies
seem to possess a lot of it. This can be checked by determining the 
dynamical masses of TDGs.

However, the existence of the most massive Tidal Dwarf Galaxies or even
of their gaseous progenitors as independent entities has been challenged
(see Hibbard et al., in this volume). Indeed, an apparent accumulation of tidal
material could in reality be the result of a projection effect.
In the 3D space, tidal tails are curved. Seen edge-on, they appear as 
linear structures and may present at their tip fake mass concentrations due to the
presence of projected material along the line of sight.
Using numerical simulations, we have shown that such projection effects 
 have a kinematical signature (Bournaud et al. 2004). Indeed,
as shown in Fig.~1, the large-scale velocity gradient 
corresponding to the streaming motions along the expanding tails
 changes its sign before their apparent extremity whenever a projection
 effect exists. 
%In case of a very extended and highly curved tail, a loop
%in the position-velocity diagram of this tail seen edge-on, is expected.

\begin{figure}[ht!]
\centering
\includegraphics[width=7cm]{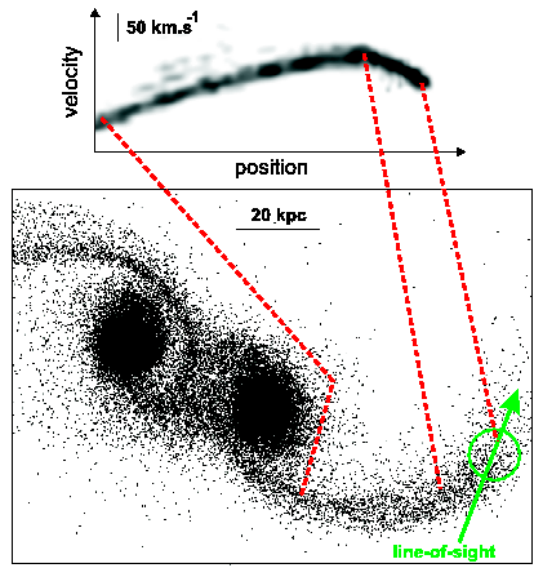} \caption{Numerical simulations  of
 two interacting galaxies. The kinematics of one of the tidal tails is analyzed as if 
it were observed edge-on, assuming that the line-of-sight is aligned with its extremity.
 This projection effect may result in an apparent accumulation of matter at the tip 
of the observed tail. The position-velocity diagram derived from this simulation shows 
a velocity gradient, mainly related to streaming motions along the tail. The sign of this 
gradient changes before the extremity of the tidal tail, a result which is obtained 
whenever a part of the tail is aligned with the line-of-sight. Conversely, a velocity 
gradient that has a constant sign rules out the presence of a projection effect.
 In addition to the large-scale motions, each TDG or gravitational clump can have 
an inner velocity gradient, related to its own dynamics.} \label{simu}
\end{figure}

Such a study requires observations with an instrument enabling
the access to both the large-scale and small-scale dynamical structures. 
 Synthesis arrays of radiotelescopes allow to probe the largest
ones but do not have enough sensitivity to
reach the $1\arcsec$ spatial resolution required to sample
correctly the nearby TDGs progenitors. Because of their low surface brightness, a direct study 
of the stellar kinematics of tidal features is at the limits of today's telescopes and
detectors. The internal dynamics can be more easily 
approached in the ionized gas component through spectroscopic
observations of the emission lines. Slit '1D' spectroscopy already
provides some information (see Weilbacher et al. 2003) but is largely insufficient given the
complex morphology of colliding galaxies. The Fabry-Perot technique appears 
as an ideal tool as it combines an integral-field capability, a high spatial and spectral resolution
and a large field of view. We present here Fabry-Perot
observations of several interacting systems where TDG candidates had been
previously identified. Observations were carried out at the
European Southern Observatory 3.6m telescope and at the
Canada-France-Hawaii 3.6m telescope. The pixel size on the sky
varies from 0.86 to 0.91 arcsec; the FOV from 170''$\times$170''
to 440''$\times $440'' and the velocity sampling from 10 to 16
km.s$^{-1}$. The data reduction procedure has been extensively
described (Amram et al. 1998 and references therein).

\section{Fabry-Perot observations of interacting galaxies.}
\noindent We have observed systems that present  HII regions located all 
along the tidal tails, which ensures a fair sampling of the kinematics of the tidal 
structures. The most active and massive HII regions generally  stand at their 
extremities, and correspond to accumulations of HI with typical masses 
of $10^9$ M$_{\odot}$ , and are thus massive TDG candidates.
Two kind of kinematical features are observed in the tidal debris:\\
-- large--scale velocity gradients resulting from streaming motions and the
projection of velocities along the line-of-sight\\
-- small--scale  velocity gradients, possibly related to the collapse or 
the rotation of decoupled objects in the tails.

\subsection*{Streaming motions and projection effects}
%We have applied the kinematics--based  criteria described in the introduction  to 
%investigate the projection effects along tidal tails. 
We present here our results for three
 systems (more in Bournaud et al., 2004) where the presence of massive TDGs
had previously been claimed: IC 1182, a merger in the Hercules cluster
 (see  Fig.~2 and  van Driel et al., in these proceedings), Arp 105 (NGC 3561, ``The guitar"), 
an interacting system between a spiral and an elliptical
(see Fig.~3) and Arp~242 (NGC 4676, ``The Mice"), an interacting system between two spirals
with prominent tidal tails (see Fig.~4, and Hibbard et al., in this volume).

\begin{figure}[ht!]
\centering
\includegraphics[width=8.25cm]{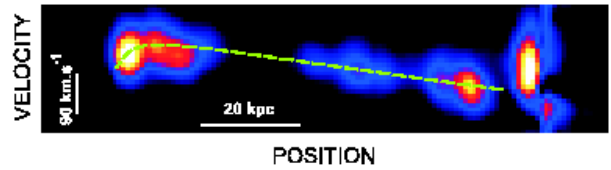} \caption{\textbf{IC~1182}:
Position-velocity diagram along the western tidal tail (see van Driel et al.  in
 these proceedings).  Several HII regions are present along the
 tidal tail, the most intense one being at its end.  The velocity gradient changes its sign before 
the massive TDG candidate, which is consistent  with the kinematical signature of a projection 
effect (see Fig.~1).}
\end{figure}

The tail of one system, IC 1182,  shows a large-scale kinematics  suggesting that
 projection effects may play a role, even though the presence of a genuine 
accumulation of matter, corresponding to a TDG, cannot be excluded. 

\begin{figure}[ht!]
\centering
\includegraphics[width=8.25cm]{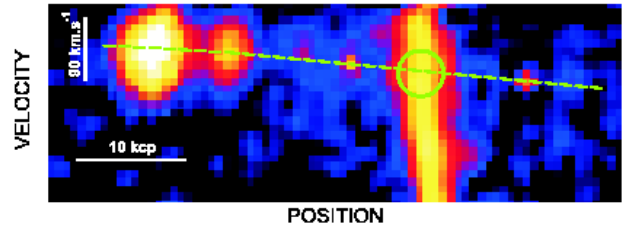} \caption{\textbf{Arp~105}:
Position-velocity of the southern tidal tail of Arp~105. The origin of the tail is to the right. 
An HII region decoupled 
from the elliptical, but hidden by it on this diagram (see Bournaud et al. 2004), is encircled. 
The large-scale velocity curve is highlighted by the dashed line. It is constant along the tail. According 
to our numerical simulations, this suggests that the massive TDG candidate located at the extremity 
is not due to a projection effect.} \label{A105}
\end{figure}

\begin{figure}[ht!]
\centering
\includegraphics[width=8.25cm]{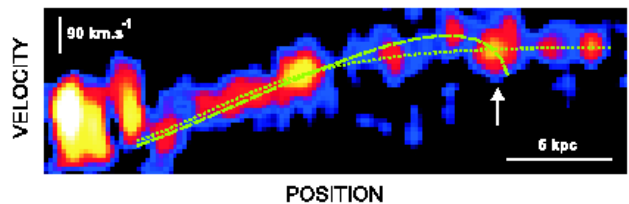} \caption{\textbf{Arp~242}:
Position-velocity diagram along the northern tidal tail. The origin of the tail is to the left. The position 
of one massive TDG candidate is indicated by the arrow.  The velocity curve may be similar to
that expected for a model with a projection effect, i.e. having  a change in the sign of the velocity 
gradient (long-dashed line). However, the kinematics of the last 5 kpc is then not fitted at all. 
A model with a  monotonically increasing velocity  (short-dashed line), without any projection
effect,  seems more robust.}
\end{figure}

The shapes of the  position-velocity diagrams of the two other systems, Arp~105 and Arp~242 
rule out  projection effects: the massive condensations observed at the extremity of their
tidal tails seem real. This is consistent with the detection in these
TDG candidates of molecular gas probably formed in the dense HI tidal clouds (Braine et al., 2001).

%The result of our simulations can also be applied to other systems ; for instance they 
%can prove that the massive TDG candidate in the eastern tail of NGC~7252 (``Atoms for Peace'') is 
%a genuine object (see details in Bournaud et al. 2004).

\subsection*{Inner dynamics of TDGs}

Beside providing information on the large-scale kinematics of tidal tails, our FP observations 
were useful to study the internal small-scale kinematics of the condensations present along 
the tidal tails. Velocity gradients of typically 50--100 km/s were observed  in the tails of 
Arp~105, IC~1182, and NGC~5291. We best explained them by the local effect of gravity. Because 
of the higher spectral resolution of the FP instrument,  the  gradients are generally smaller than those 
initially measured in earlier observations with  long slit spectroscopy.

\begin{figure}[ht!]
\centering
\plottwo{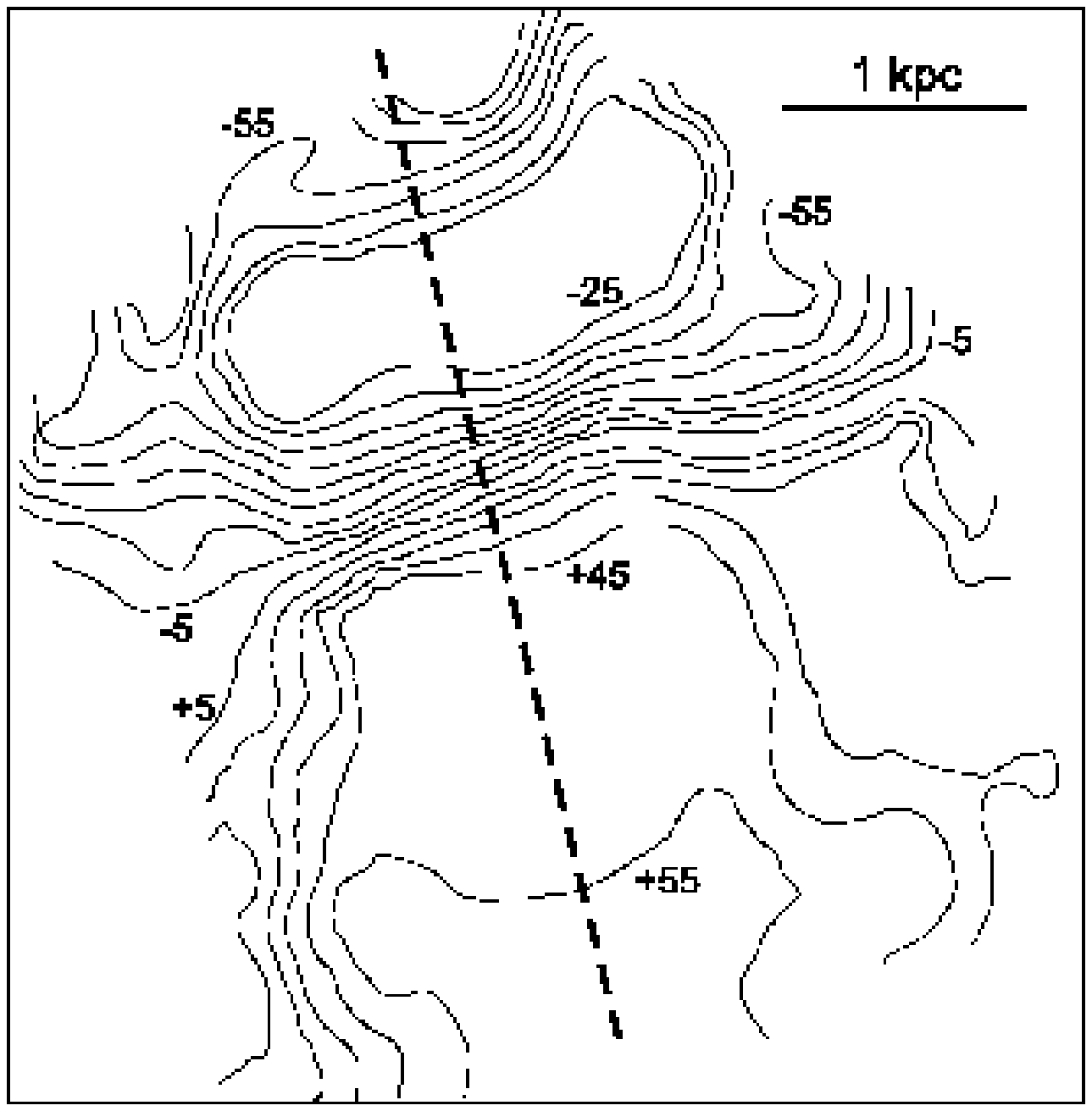}{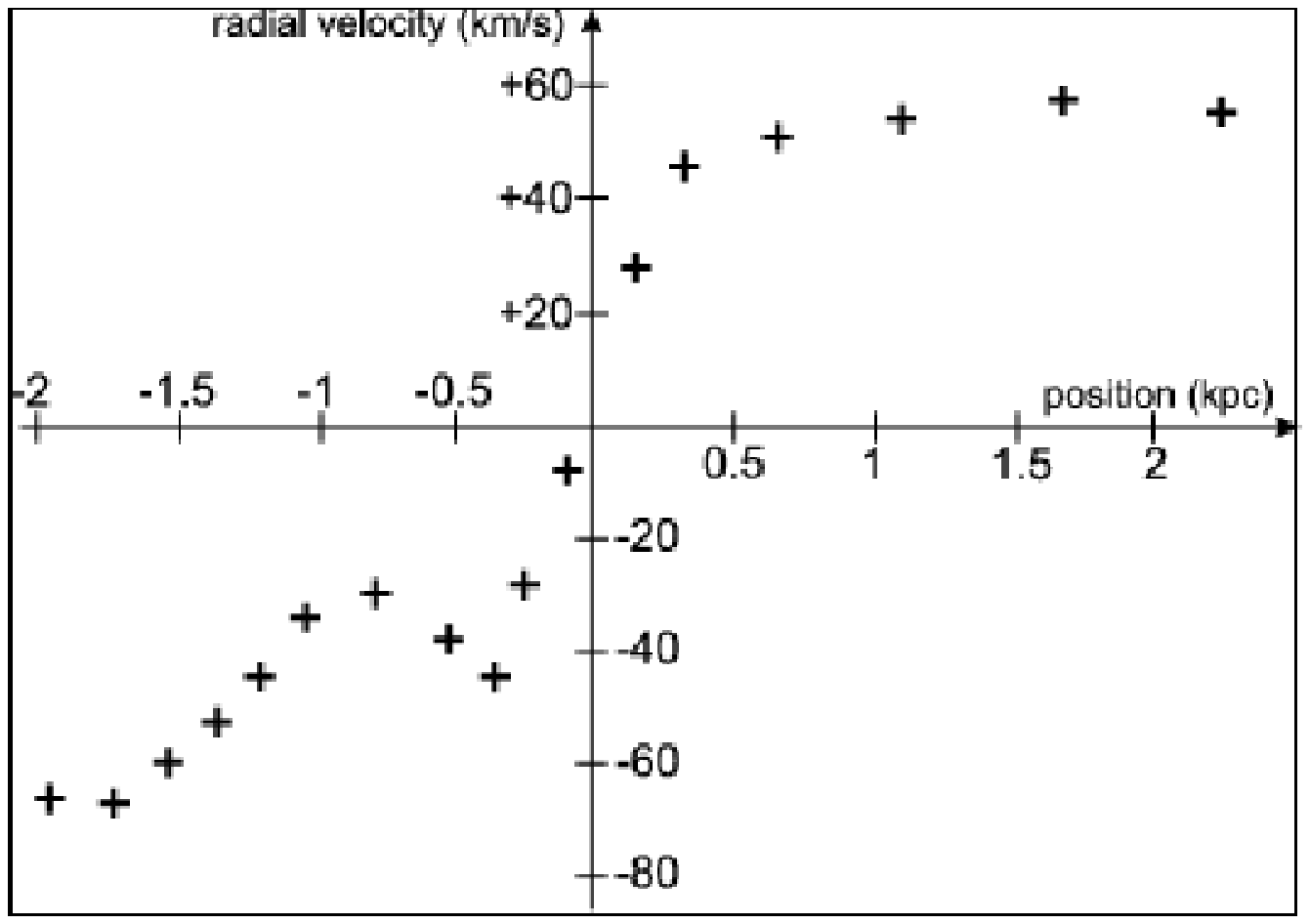}
\caption{\textbf{NGC~5291}: Left: Isovelocities diagram 
of the northern TDG candidate. Labels are in km.s$^{-1}$. This diagram strongly suggests that the system is
already rotating, even though other motions disturb the classical spider shape. This TDG candidate 
is thus decoupled from the large-scale HI structures and its inner dynamics can be studied. Right: a velocity 
gradient of 100 km.s$^{-1}$, with a disturbed rotation curve, is observed along the dashed line.}
\end{figure}

The more striking object  showing an inner velocity gradients is observed in the northern structures 
of NGC~5291 (``The SeaShell"). This system is remarkable for the extremely large and massive 
HI ring surrounding a disturbed lenticular galaxy. Numerous intergalactic HII regions are located
towards the HI clumps in the ring (Duc \& Mirabel 1998)\footnote{Although the origin of the ring is 
strictly speaking probably not tidal, it was undoubtedly shaped by a past collision,
and the small-scale phenomena that occurred in the collisional debris, in particular
the onset of star formation, are probably the same as in more typical Tidal Dwarf Galaxies.}.  
 The northern TDG candidate shows a velocity 
gradient of 100 km.s$^{-1}$ throughout a region of 2.4 kpc, and the isovelocity diagram is that of a rotating
 but disturbed system (see Fig.~5). If we assume an inclination of approximatively 45 degrees (based on the
 isovelocities diagram and on the shape of the outer H$_{\alpha}$ isophotes), the deprojected velocity 
gradient is 140 km.s$^{-1}$. Assuming  that the TDG is a relaxed rotating system, we  find a 
dynamical mass of 1.4 10$^{9}$ M$_{\odot}$. We estimate that the contribution due to non--circular motions 
(that are visible in the isovelocity diagram) is of the
same order (Braine et al., 2001).  The HI mass of the TDG is 2.6 10$^{9}$ M$_{\odot}$.
 Taking into account the additional contribution of the molecular gas and the stars, the visible 
mass matches the dynamical mass.  A straightforward conclusion would be that this system contains
no dark matter, as already suggested by Duc \& Mirabel (1998). However,  because of
our limited spectral resolution,  the measure of the dynamical mass 
%provides a lower limit for the total mass. On the other hand, 
is still uncertain and the HI mass may be an upper limit if the HI beam is larger 
than the clump of HII regions. It is thus not
impossible that the  visible mass is  lower than the dynamical mass and that this
TDG  contains some dark matter. 

\section{Summary and conclusions}
\noindent 
We have shown that the  kinematics of tidal tails may be used
to check whether the claimed massive TDG candidates observed at their
tip  are real bound entities or the result of projection effects. With their
high spatial resolution, three-dimensional  Fabry-Perot  observations in the H$\alpha$ line turn out to 
be well adapted to the problem. 
Analyzing position-velocity diagrams derived from FP data obtained at the CFHT and 
at the ESO 3.6m, we were able to  rule our projection effects as the main contributor to the
condensations seen in the northern tail of Arp~242 (NGC 4676) and in 
the southern tail of Arp~105.  We cannot  exclude that they play a role in the merger
 IC~1182. 
The existence of massive entities observed near the tip 
of several tidal tails is actually supported by the numerical simulations
of interacting galaxies by Bournaud, Duc \& Masset (2003) in which
tidal objects with comparable masses are formed.
% (see also Duc et al, in these proceedings).

We have also studied the inner kinematics of a few TDG candidates. 
Several velocity gradients were found. The kinematics of the northern mass concentration of NGC~5291 
is consistent with that expected for a  self-gravitating rotating  object. Its inferred dynamical and 
luminous mass compare, but higher resolution observations are required to prove that the object does 
not contain a certain amount of  dark matter.

\end{document}